\documentclass[12pt]{article}
\def\timesbox{\hbox{$\scriptscriptstyle\times$}}
\def\ant{ {{\lower 1ex  \timesbox} \atop {\raise 1.5ex  \timesbox}}}
\newcommand{\Zop}{{\hbox{ Z\kern-1.6mm Z}}}\begin{scriptsize}\end{scriptsize}
\newcommand{\beq}{\begin{equation}}
\newcommand{\eeq}{\end{equation}}
\newcommand{\bea}{\begin{eqnarray}}
\newcommand{\eea}{\end{eqnarray}}
\newcommand{\ra}{\rangle}
\newcommand{\la}{\langle}
\newcommand{\lt}{\left}
\newcommand{\rt}{\right}
\newcommand{\Iop}{\relax{\rm I\kern-.18em I}}
\newcommand{\one}{{\hbox{ 1\kern-1.2mm l}}}
\newcommand{\T}{{\cal T}}

\newcommand{\dt}{\delta}
\newcommand{\del}{\partial}

\newcommand{\D}{{\cal D}}
\newcommand{\eps}{\epsilon}

\newcommand{\lam}{\lambda}

\newcommand{\s}{\sigma}

\newcommand{\xh}{\hat x}
\newcommand{\ph}{\hat p}
\newcommand{\pih}{\hat \pi}

\begin{document}

{}~
{}~
\hfill\vbox{\hbox{IMSc/2009/12/15}}
\break

\vskip 2cm

\centerline{\Large \bf On a coordinate independent description}
\centerline{\Large \bf of string worldsheet theory}

\medskip

\vspace*{4.0ex}

\centerline{\large \rm Partha Mukhopadhyay }

\vspace*{4.0ex}

\centerline{\large \it The Institute of Mathematical Sciences}
\centerline{\large \it C.I.T. Campus, Taramani}
\centerline{\large \it Chennai 600113, India}

\medskip

\centerline{E-mail: parthamu@imsc.res.in}

\vspace*{5.0ex}

\centerline{\bf Abstract}
\bigskip

We study worldsheet conformal invariance for bosonic string
propagating in a curved background using the hamiltonian formalism. In
order to formulate the problem in a background independent manner we
first rewrite the worldsheet theory in a language where it describes a
single particle moving in an infinite-dimensional curved
spacetime. This language is developed at a formal level without
regularizing the infinite-dimensional traces. Then we adopt DeWitt's
(Phys.Rev.85:653-661,1952) coordinate independent formulation of
quantum mechanics in the present context.  Given the expressions for the classical Virasoro generators, this procedure enables us to define the coordinate invariant quantum analogues which we call DeWitt-Virasoro generators. This framework also enables us to calculate the invariant matrix elements of an arbitrary
operator constructed out of the DeWitt-Virasoro generators between two
arbitrary scalar states. Using these tools we further calculate the DeWitt-Virasoro algebra in spin-zero representation. The result is given  by the Witt algebra with additional anomalous terms that vanish for Ricci-flat backgrounds.  
Further analysis need to be performed in order to precisely relate this with the beta
function computation of Friedan and others. Finally, we explain how this analysis improves the understanding of showing conformal invariance for certain pp-wave that has been recently discussed using hamiltonian framework.

\newpage

\tableofcontents

\baselineskip=18pt

\section{Introduction and summary}
\label{s:intro}

Equations of motion (EOM) for backgrounds in string theory are derived
from the condition of worldsheet conformal invariance\footnote{Some of
  the original references are 
  \cite{friedan80, lovelace84, fradkin85, sen85, callan85, fridling86}.}. Although this condition has mostly
been studied by computing the beta functions of the nonlinear sigma
model using background field method \cite{friedan80}, a BRST hamiltonian approach has also
been discussed in the literature \cite{maharana87, akhoury87, das87,
  fubini89, alam89, diakonou90, wess90} (See also \cite{jain87}). An important issue in
the latter is to find out how to define the constraint generators at
the quantum level. This problem has been dealt with by considering a weak field
approximation near flat space and/or using worldsheet supersymmetry.
 
Recently a similar method has been used \cite{kazama08, pm08, pm0902,
  pm0907} to study exact conformal invariance of the worldsheet theory
in type IIB R-R plane-wave background \cite{blau01} using
Green-Schwarz formalism in semi-light-cone gauge
\cite{berkovits04}. In this case neither we are close to flat space
nor do we have worldsheet supersymmetry. However, knowing the exact string
spectrum through light-cone gauge analysis \cite{metsaev01} and some special properties
of the background help us to fix the quantum definition of the
energy-momentum (EM) tensor\footnote{Authors in \cite{kazama08} argued
with certain analysis that the relevant worldsheet theory should not have a smooth flat
space limit and suggested that the complete EM tensor be defined with an unusual
operator ordering, called phase-space normal ordering (similar ordering was
considered earlier in the literature in \cite{wess90}). 
However,
it was pointed out in \cite{pm08} using the universality 
argument of \cite{universality} that such a limit should be smooth
generically for any pp-wave and the definition of \cite{kazama08}
leads to a spectrum of negative conformal dimensions. An alternative definition was
suggested in \cite{pm0902} where such an ordering is applied only to the
interaction part. It was argued that this definition leads to the correct physical
spectrum.}.    

The next issue is to understand how to compute Virasoro anomaly. One
may naively think that such anomaly terms can be computed by directly
calculating the commutators among the right and left moving EM tensor
components using the basic canonical commutation relations. However,
it was shown in \cite{pm0902} that this direct method leads to results
that suffer from operator ordering ambiguity. It did not seem to be an
issue in the earlier 
works as all the composite operators were ordered according to the
normal ordering relevant to flat space. However, in our case the exact
vacuum is not close to the flat space vacuum and it is 
{\it a priori} not clear which ordering needs to be considered. A
  supersymmetry  argument has been used in \cite{pm0907} to compute
  certain integrated forms of the anomaly terms indirectly. Such
  analysis works because there are certain relations among the
  supercurrents and the EM tensor components which hold true in the
  non-perturbative vacuum and are enough to establish Virasoro
  algebra\footnote{This is analogous to the case of NSR
  superstrings where one uses superconformal algebra. See for example \cite{fubini89}.}. However, the
method itself works only for on-shell backgrounds which are supersymmetric.    

Given the above discussion one may wonder if there exists a notion of Virasoro algebra 
in the quantum theory which can be discussed in a vacuum independent way. This question
seems natural in the context of a background independent formulation as a vacuum state 
has information about a particular background. In ordinary quantum mechanics this 
question is not any special as one can indeed define all the operators of the theory 
before solving for the energy eigenstates. The difference is that here we have 
infinite number of degrees of freedom and if we sum all their contributions naively we
get divergences. In this work we will still try to formulate the problem in a background 
independent manner in the sense described above by going to a particle-like 
description. We will find certain interesting result and it will be
interesting to explore further along this direction. Below we 
describe our construction and result in more detail and explain
why the particle-like description is well-suited for our purpose.

At the classical level the new description is obtained by re-writing the worldsheet 
theory in a language where it describes a particle moving in an infinite-dimensional 
curved spacetime (subject to certain potential). This way the infinite number of 
degrees of freedom of the string is given an interpretation of number of spacetime dimensions. We will 
define the background independent version of the quantum Virasoro generators in this new language. In general the expressions 
will have divergent terms. However, all such divergences will appear
as infinite-dimensional traces and therefore will be formally
tractable. A suitable regularization procedure needs to be developed
to cure such divergences. We will leave this for future work.  
 
To describe the quantum mechanics in a manifestly covariant manner we
follow the argument of DeWitt in \cite{dewitt52} (see also
\cite{dewitt57, omote72})\footnote{DeWitt's formulation was applied to
  string theory earlier by Lawrence and Martinec in
  \cite{lawrence95}.} and the particle-like description will provide a
natural framework to adopt this idea. In \cite{dewitt52} DeWitt identified
general coordinate transformations (GCT) with a subgroup of all
unitary transformations in quantum mechanics. The method shows how to
construct the quantum analogs of dynamical 
quantities in terms of the general phase-space coordinates such that the expectation values of such operators between two arbitrary states are given by covariant expressions in position space representation. 

There are a few generalizations involved in our work from
DeWitt's original work. The analysis of
\cite{dewitt52} considered a non-relativistic particle so that the
general covariance was sought only for the spatial slice. In our case
we adopt the infinite-dimensional language for the matter part of the
worldsheet theory in conformal gauge. The resulting particle-like theory
looks like a worldline theory with full covariance in spacetime. A
more important difference is having infinite number of Virasoro
generators instead of only the hamiltonian as in DeWitt's case. 
Because of the presence of infinite number of dimensions the theory possesses 
certain {\it shift} properties which look unusual from the particle
point of view. These properties dictate the behavior of the theory
under certain shift of the spacetime dimensions, i.e. the string
modes. Since Virasoro generators relate different string modes, such
shift properties are inherently related to the existence of these
generators. 

The quantum generators defined in the sense of DeWitt, hereafter
called {\it DeWitt-Virasoro generators}, are ordered in a particular
way and are designed to produce covariant results in spin-zero representation. Given this background independent definition, we next
proceed to compute the algebra, hereafter called {\it DeWitt-Virasoro
  algebra}, satisfied by these generators in spin-zero representation. The result is 
given by the Witt algebra with   additional anomalous terms that vanish for Ricci-flat backgrounds. Notice that the central charge terms of the Virasoro algebra do not appear in this result. Being constructed in a background independent way, the DeWitt-Virasoro generators are not normal ordered with respect to any particular vacuum. It is expected that the actual quantum Virasoro generators for a specific conformal background can be obtained by first specializing to that background and then normal ordering the DeWitt-Virasoro generators with respect to the relevant vacuum. 
The central charge terms are expected to arise in the algebra of such quantum Virasoro generators. This procedure is understood for flat and the pp-wave background discussed in \cite{pm0902}. Understanding of this in a generic sense is an important question for our construction.

Finally, as an application of the present background independent framework, we discuss how it explains conformal invariance for the pp-wave considered in \cite{kazama08, pm08, pm0902}. As mentioned earlier, the anomaly terms were shown in \cite{pm0902} to suffer from operator ordering ambiguity. If these terms are ordered according to the phase-space normal ordering then the correct EOM is reproduced. However, the justification of such ordering prescription was not clear. Here we argue that the computation of \cite{pm0902} can be viewed as a special
case of our present analysis. Therefore the anomaly terms computed in that work are same as the DeWitt-Virasoro anomaly terms discussed here. Our present result suggests that the terms computed in \cite{pm0902} should vanish. However, this also implies that the Ricci-term found here, which gives the EOM for the background, was missing in the computation of \cite{pm0902}. This apparent discrepancy may be resolved by the
following observation. The Ricci-term that we obtain here involves certain contractions with the Ricci tensor. The whole term vanishes, though the Ricci tensor itself does not, for the relevant pp-wave because of its special properties.   

Given the above results, there are a number of technical issues which
deserve further attention. For example, in our computation the Ricci-flatness
condition comes from terms that arise only in the left-right
commutation relations. From the present analysis it is not clear why
this is so. A better understanding of this with the inclusion of other
massless fields in the background is desirable. In particular, it will
be interesting to investigate if the {\it DeWitt-Virasoro anomaly}
terms contain the infinite-dimensional analogue of the low energy EOM
for the massless fields as obtained in the beta function
computation. Superstring extension of this result will also be
interesting to find using pure spinor
formalism \cite{purespinor}. It should also be investigated how the present
analysis may be extended to higher spin representations. The main
conceptual issue involved in our work is to understand how to
interpret the current framework in more conventional terms. For example, how the Ricci-flatness condition found here should be related to the usual beta function computation of Friedan \cite{friedan80}. We hope to come back to these questions in future. 

The rest of the paper is organized as follows. The infinite-dimensional language is explained in  sec.\ref{s:map}. The construction of the DeWitt-Virasoro generators has been discussed in sec.\ref{s:vir-gen}. We summarize the results for the DeWitt-Virasoro algebra in sec.\ref{s:vir-alg}. We discuss the flat and the pp-wave backgrounds as special cases of the present construction in sec.\ref{s:special}. Some of the technical derivations are presented in a few appendices.

\section{Mapping to infinite dimensions}
\label{s:map}

We consider a bosonic closed string propagating in a $D$ dimensional curved background, hereafter 
called the physical spacetime, with metric $G_{\mu \nu}$. We work in the 
conformal gauge of the worldsheet theory so that the ghosts are given by the standard $(b,c)$ 
systems. For the purpose of the present work we will be concerned only with the matter part of 
the theory. The relevant classical lagrangian is given by,
\bea
L &=& {1\over 2} \oint {d\s \over 2\pi} ~G_{\mu \nu}(X(\s))\lt[\dot X^{\mu}(\s) \dot X^{\nu}(\s) 
- \del X^{\mu}(\s) \del X^{\nu}(\s)\rt]~,
\label{Lworldsheet}
\eea
where $\oint \equiv \int_0^{2\pi}$, $\mu = 0,1, \cdots , D-1$. A dot and a $\del$ denote 
derivatives with respect to worldsheet time-coordinate $\tau$ and space-coordinate $\s$ 
respectively. 
We recast this lagrangian in a form that describes a single particle moving in an 
infinite-dimensional curved spacetime subject to certain potential,
\bea
L(x, \dot x) &=& {1\over 2} g_{ij}(x) \lt[\dot x^i \dot x^j -  a^i(x) a^j(x) \rt]~,
\label{Linfinite}
\eea
where $x^i$ are the general coordinates of the infinite-dimensional
spacetime. The index $i$ is given by an ordered pair of indices,
\bea
i = \{\mu, m \}~,
\eea
where $m \in Z$ is the string-mode-number such that\footnote{Throughout 
the paper we will make the following type of index identifications: $i=\{\mu, 
m \}$, $j=\{\nu, n \}$, $k=\{\kappa, q\}$. \label{index}},
\bea
x^i &=& \oint {d\s \over 2\pi}~ X^{\mu}(\s) e^{-im\s} ~,\cr
g_{ij}(x) &=& \oint {d\s \over 2\pi} ~G_{\mu \nu}(X(\s))e^{i(m+n)\s}~, \cr
a^i(x) &=&  \oint {d\s \over 2\pi}~ \del X^{\mu}(\s) e^{-im\s}~.
\label{xga-XGdX}
\eea

As mentioned in the previous section, we will mainly work using the
infinite-dimensional language. Below we discuss certain
properties of this language that will be relevant for our study.
\begin{enumerate}
\item
We have claimed that the worldsheet theory (\ref{Lworldsheet}) has an interpretation to
be generally covariant in the infinite-dimensional sense. To see this
explicitly let us consider a GCT in the physical spacetime: $X^{\mu}\to X'^{\mu}(X)$ with transition 
function $\Lambda^{\mu}_{~\nu}(X) = {\del X'^{\mu} \over \del X^{\nu}}$ and its inverse 
$\Lambda_{\mu}^{~\nu}(X) = {\del X^{\nu} \over \del X'^{\mu}}$. This induces a GCT in the 
infinite-dimensional spacetime: $x^i \to x'^i$ such that the Jacobian matrix  
$\lambda^i_{~j}(x)={\del x'^i \over \del x^j}$ and its inverse $\lambda_i^{~j}(x)={\del x^j\over 
\del x'^i}$ are given by,
\bea
\lambda^i_{~j}(x) = \oint {d\s \over 2\pi}~ \Lambda^{\mu}_{~\nu}(X(\s)) e^{i(n-m)\s}~, \quad 
\lambda_i^{~j}(x) = \oint {d\s \over 2\pi}~ \Lambda_{\mu}^{~\nu}(X(\s)) e^{i(m-n)\s}~.
\eea
One can then show (see appendix \ref{a:map}) that $g_{ij}(x)$ and $a^i(x)$ transform as tensors, 
\bea
g'_{ij}(x') = \lambda_i^{~k}(x) \lambda_j^{~k'}(x) g_{kk'}(x)~, \quad
a'^i(x') = \lambda^i_{~j}(x) a^j(x)~.
\label{GCT-g}
\eea

\item
Using the map in (\ref{xga-XGdX}) one can relate any field in the infinite-dimensional spacetime 
constructed out of the metric, its inverse, $a^i(x)$ and their derivatives to a non-local 
worldsheet operator. A class of examples, which will prove to be useful for us, is given by a 
multi-indexed object $u^{i_1j_1\cdots}_{i_2j_2\cdots}(x)$ constructed out of the metric, its 
inverse, their derivatives and $a^i(x)$ (but not its derivatives) such that 
$u^{i_1j_1\cdots}_{i_2j_2\cdots}(x)$ can not be factored into pieces which are not contracted 
with each other. In this case one can construct a local worldsheet operator
 $U^{\mu_1 \nu_1 \cdots}_{\mu_2 \nu_2 \cdots}(X(\s))$ simply 
by performing the following replacements in the expression of 
$u^{i_1j_1\cdots}_{i_2j_2\cdots}(x)$:
\bea
g_{ij}(x) \to G_{\mu \nu}(X(\s))~, \quad g^{ij}(x) \to G^{\mu \nu}(X(\s))~,\quad \del_i \to 
\del_{\mu}~, \quad a^i(x) \to \del X^{\mu}(\s)~.
\label{replace}
\eea
The two objects $u^{i_1j_1\cdots}_{i_2j_2\cdots}(x)$ and $U^{\mu_1 \nu_1 \cdots}_{\mu_2 \nu_2 
\cdots}(X(\s))$ are related to each other by the following general rule (see appendix 
\ref{a:map}):
\bea
u^{i_1j_1\cdots}_{i_2j_2\cdots}(x) \sim [2\pi \delta (0)]^N \oint {d\s \over 2\pi} ~U^{\mu_1 
\nu_1 \cdots}_{\mu_2 \nu_2 \cdots}(X(\s)) e^{i(m_2+n_2+\cdots)\s -i(m_1+n_1+ \cdots)\s}~,
\label{th-rule}
\eea
where $N$ is the number of traces in $u$ and the argument of the Dirac delta function $\dt (0)$ 
appearing on the right hand side is the worldsheet space direction:
\bea
\dt(0) = \lim_{\s \to \s'} \dt (\s-\s') =\lim_{\s\to \s'} {1\over 2\pi} \sum_{n\in 
Z}e^{in(\s-\s')}~.
\label{delta-0}
\eea
The way one gets $N$ factors of $\dt(0)$ on the right hand side is as follows:
Each infinite-dimensional trace breaks up into a trace in the physical spacetime which appears in 
the expression of $U$, and a sum over all the string modes which gives rise to a factor of 
$\displaystyle{\sum_{n\in Z} 1 =2\pi \dt(0)}$. We relate the two sides of (\ref{th-rule}) by the 
symbol $\sim$ to indicate that such a manipulation is understood only at a formal level. The 
relation (\ref{th-rule}) implies that $u$ enjoys the same tensorial 
properties in the infinite-dimensional spacetime as $U$ does in the physical spacetime (provided 
$g_{ij}$ and $a^i$ have the right tensorial property, which is indeed the case as we have already 
discussed).

\item
In the infinite-dimensional language the problem at hand possesses certain shift properties which 
can be written as:
\bea
u^{i_1+i i_2 \cdots}_{j_1j_2\cdots} &=& u^{i_1 i_2+i \cdots}_{j_1j_2\cdots} = u^{i_1 i_2 
\cdots}_{j_1-i j_2\cdots} = u^{i_1 i_2 \cdots}_{j_1j_2-i\cdots} = \cdots ~, \cr
\del_{j+l} a^{k+l}(x) &=& \del_j a^k(x)+ i(l) \delta^k_j~,
\label{shift}
\eea
where the factor of $i$ in the second term of the last equation is the imaginary number.
Given the spacetime index $i$ as in footnote \ref{index}, we have defined $(i)=m$. A shift
in the infinite-dimensional index is defined to be $i+j=\{\mu, m+n\}$\footnote{Notice that we choose the physical spacetime index corresponding to $i+j$ by the one associated with the first index (i.e. $i$) appearing in the shift. We will follow this convention in all our expressions.}. It is now obvious that the first relation 
of (\ref{shift}) is a direct consequence of (\ref{th-rule}). The second relation can be obtained 
from the following one:
\bea
\del_ja^k = i(j) \delta^k_j~.
\label{del-a}
\eea
The easiest way to get this is to notice that the definitions in (\ref{xga-XGdX}) imply that the 
infinite dimensional model in (\ref{Linfinite}) corresponds to the string worldsheet theory only 
for the linear profile $a^k(x) = (k) x^k$. Alternatively, one can directly calculate the left 
hand side of (\ref{del-a}) using the third equation in (\ref{xga-XGdX}) and eq.(\ref{x-X}). The 
result is given by $in\delta_{n,q}\delta^{\kappa}_{\nu}=i(j)\dt_j^k$.

\end{enumerate}
\section{DeWitt-Virasoro generators}
\label{s:vir-gen}

The goal of this section is to arrive at the background independent version of the quantum Virasoro generators.
We will start with the standard expressions for the classical EM tensor and write the classical Virasoro generators in the infinite-dimensional language. Then after quantizing the system we will use DeWitt's argument to define the quantum DeWitt-Virasoro generators.

The right and left moving components of the classical EM tensor are given by,
\bea
\T(\s) &=&{1\over 4} \lt( K(\s)-Z(\s)+V(\s) \rt) = \sum_{m \in Z}L_m e^{im\s}~, \cr
\tilde \T(\s) &=&  {1\over 4} \lt( K(\s)+Z(\s)+V(\s) \rt) = \sum_{m\in Z} \tilde L_m e^{-im\s}~,
\eea
respectively, where,
\bea
K(\s) &=& G^{\mu \nu}(X(\s))P_{\mu}(\s)P_{\nu}(\s) = \sum_{m\in Z} K_m e^{im\s}~, \cr
Z(\s) &=& 2 \del X^{\mu}(\s) P_{\mu}(\s) = \sum_{m\in Z} Z_m e^{im\s}~, \cr
V(\s) &=& G_{\mu \nu}(X(\s)) \del X^{\mu}(\s) \del X^{\nu}(\s) = \sum_{m\in Z} V_m e^{im\s}~.
\eea
The conjugate momentum is given by: $P_{\mu}=G_{\mu\nu}(X)\dot
X^{\nu}$. It is related to the momentum in the infinite-dimensional
language, i.e. $p_i=g_{ij}(x)\dot x^j$ according to the  
same rule in (\ref{th-rule}). The classical Virasoro generators $L_m$ and $\tilde L_m$ can now be expressed 
in terms of the Fourier modes $K_m$, $Z_m$ and $V_m$, which can, in turn, be expressed in the 
infinite-dimensional language. The results are as follows:
\bea
4 L_{(i)} = K_{(i)}-Z_{(i)}+V_{(i)} ~, \quad
4 \tilde L_{(i)} = K_{(\bar i)}+Z_{(\bar i)}+V_{(\bar i)}~,
\label{LLtilde-classical}
\eea
where we have defined $\bar i=\{\mu, -m\}$ and,
\bea
K_{(i)} &=& g^{k l+i}(x) p_k p_l~, \quad Z_{(i)} =2 a^{k+i}(x) p_k~, \quad V_{(i)}=g_{kl}(x) 
a^k(x) a^{l+i}(x)~.
\label{KZV-classical}
\eea
Notice that a Virasoro generator is a scalar, but has a string mode index $(i)=m$ which, in the 
infinite-dimensional language, appears to be a shift of the spacetime index, as evident from 
eqs.(\ref{KZV-classical}).

Poisson brackets of the generators in (\ref{LLtilde-classical}) should satisfy the classical 
Virasoro algebra. Notice that since GCT is a canonical transformation which preserves the Poisson 
brackets, it should be possible to write such brackets in a manifestly covariant manner. We have 
derived these brackets in the infinite-dimensional language in appendix \ref{a:classical}.

We now quantize the system:
\bea
[\xh^i, \ph_j] = i \alpha' \delta^i_j ~.
\label{can-comm}
\eea
We work in the Schr$\ddot{\rm{o}}$dinger picture so that the operators
do not have explicit $\tau$ dependence. 
The idea is to keep the general covariance manifest in the quantum theory. GCT of any operator independent of momenta does not suffer from any ordering ambiguity and therefore straightforward 
to compute. Transformation of the momentum operator, which preserves the canonical commutation relations, is taken to be
\cite{dewitt52, dewitt57, omote72}:
\bea
\ph_i \to \ph'_i = {1\over 2} (\lam_i^{~j}(\xh) \ph_j + \ph_j \lam_i^{~j}(\xh) )~.
\label{GCT-ph}
\eea
This defines GCT of an arbitrary operator constructed out of the
phase-space variables.

Let us now introduce the position eigenbasis $|x\ra$. The orthonormality and completeness 
conditions read:
\bea
\la x|x'\ra = \delta(x,x')=g^{-1/2}(x) \delta(x-x')~, \quad \int dw ~|x\ra \la x| =1~,
\label{ortho}
\eea
where $\delta (x-x')$ is the Dirac delta function, $dw = dx g^{1/2}(x)$ and $g(x) =|\det 
g_{ij}(x)|$. The position space representation of the momentum
operator is given by \cite{dewitt52},
\bea
\la x|\ph_i|x'\ra = -i\alpha' \lt[\del_i + {1\over 2} \gamma_i(x) \rt] \delta (x,x')~,
\label{p-rep}
\eea
where $\gamma_i$ are the contracted Christoffel symbols\footnote{Notice that $\gamma_i$ has a 
trace and therefore is divergent:
\bea
\gamma_i(x) = 2\pi \dt(0) \oint {d\s \over 2\pi}~ \Gamma_{\mu}(X(\s)) e^{im\s}~,
\eea
where the expression for $\Gamma_{\mu}(X)$ can be read out from (\ref{chr}) using the replacement 
(\ref{replace}).},
\bea
\gamma_j = \gamma^i_{ji}~, \quad \gamma^i_{jk} = {1\over 2} g^{il}\lt(\del_j g_{lk} + \del_k 
g_{lj} - \del_l g_{jk} \rt)~.
\label{chr}
\eea
Using $\gamma^*_i(x)=\gamma_{\bar i}(x)$ and $\del_{x^i}\delta (x,x') = - (\del_{x'^i} 
+\gamma_i(x)) \delta(x,x')$ (see \cite{dewitt52}) it is straightforward
to check that the position space  
representation in (\ref{p-rep}) is compatible with the following hermiticity properties:
\bea
(\xh^i)^{\dagger} = \xh^{\bar i}~, \quad (\ph_i)^{\dagger} = \ph_{\bar i}~.
\label{xp-dagger}
\eea

To construct the DeWitt-Virasoro generators we first define following \cite{omote72}:
\bea
\pih_j = \ph_j + {i\alpha' \over 2} \gamma_j(\xh)~, \quad \pih_j^{\star} = \ph_j - {i\alpha' 
\over 2} \gamma_j(\xh)~.
\label{pih-def}
\eea
Using (\ref{GCT-ph}) it is easy to show that these objects are transformed by left and right 
multiplications respectively under GCT:
\bea
\pih_i \to \pih'_i = \lam_i^{~j}(\xh) \pih_j~, \quad \pih^{\star}_i \to \pih'^{\star}_i = 
\pih^{\star}_j \lam_i^{~j}(\xh) ~,
\label{GCT-pih}
\eea
The quantum definition of the operators in eqs.(\ref{KZV-classical}) are given by,
\bea
\hat K_{(i)} &=& \pih^{\star}_k g^{k l+i}(\xh) \pih_l~, \cr
\hat Z_{(i)} &=& \hat Z^L_{(i)} + \hat Z^R_{(i)}~, \cr
\hat V_{(i)} &=& g_{kl}(\xh) a^k(\xh) a^{l+i}(\xh)~
\label{KZV-quantum}
\eea
where,
\bea
\hat Z^L_{(i)} = \pih^{\star}_k a^{k+i}(\xh)~, \quad \hat Z^R_{(i)} = a^{k+i}(\xh) \pih_k .
\label{ZLZR}
\eea
Given the transformation properties in (\ref{GCT-pih}), it is clear that all the operators in 
(\ref{KZV-quantum}) and (\ref{ZLZR}) are invariant under GCT and have the right classical limit. 
The left-right symmetric combination for $\hat Z_{(i)}$ considered in (\ref{KZV-quantum}) gives the right hermiticity property for the DeWitt-Virasoro generators. These operators give covariant results in the following sense.  Consider the matrix element of an arbitrary operator constructed out 
of these generators between any two scalar states. The result written in position space representation is manifestly covariant.

\section{DeWitt-Virasoro algebra}
\label{s:vir-alg}

Here we will calculate the algebra satisfied by the DeWitt-Virasoro generators $\hat L_{(i)}$ and $\hat{\tilde L}_{(i)}$ defined through eqs.(\ref{LLtilde-classical}, \ref{KZV-quantum}, \ref{ZLZR}). As mentioned earlier, for a generic background the current framework allows us to calculate this algebra only in the spin-zero representation. We have given the details of the necessary computation in appendix \ref{a:vir-alg}. The final results, written in the infinite-dimensional language, are as follows,
\bea
\la \chi| \lt\{ \begin{array}{l}
[\hat L_{(i)}, \hat L_{(j)}] = (i-j) \alpha' \hat L_{(i+j)} + \hat A^R_{(i)(j)}~,  \cr
[\hat{\tilde L}_{(i)}, \hat{\tilde L}_{(j)}] = (i-j) \alpha' \hat{\tilde L}_{(i+j)} + \hat A^L_{(i)(j)}~, \cr
[\hat L_{(i)}, \hat{\tilde L}_{(j)}] = \hat A_{(i)(j)}
\end{array} \rt\} |\psi \ra ~,
\label{scalar-alg}
\eea
where $|\chi \ra$ and $|\psi \ra$ are two arbitrary scalar states ($\tau$-dependent). The above result is the Witt algebra (without the central charge terms) with additional anomalous terms given by,
\bea
\hat A^R_{(i)(j)} &=& 0~, \cr
\hat A^L_{(i)(j)} &=& 0~, \cr
\hat A_{(i)(j)} &=& \displaystyle{{\alpha'^2\over 8} \lt(\hat \pi^{\star k+i} 
r_{kl}(\hat x) a^{l+\bar j}(\hat x) - a^{k+i}(\hat x) r_{kl}(\hat x) \hat \pi^{l+\bar j} \rt)}~,
\label{anomalies-id}
\eea 
where $r_{ij}(x)$ is the Ricci tensor in the infinite-dimensional spacetime which, according to the general map (\ref{th-rule}), is related to the same in physical spacetime, namely $R_{\mu \nu}(X)$ in the following way,
\bea
r_{ij} (x) \sim 2\pi \dt(0) \oint {d\s\over 2\pi} ~R_{\mu \nu}(X(\s)) e^{i(m+n)\s}~.
\label{rij-Rmunu}
\eea

We will now make comments on not having the central charge terms in (\ref{scalar-alg}). We have mentioned before that the DeWitt-Virasoro generators are background independent and therefore are different from the quantum Virasoro generators at a particular conformal background. Similarly, the algebra in (\ref{scalar-alg}, \ref{anomalies-id}) is not expected to be same as the quantum Virasoro algebra that occurs at the conformal point. The generic procedure of obtaining the quantum Virasoro generators and algebra at an arbitrary conformal point starting from the present construction is not understood in this work. This, however, is understood for flat space and the pp-wave background considered in \cite{pm0902} as discussed below.

\section{Flat and pp-wave backgrounds as special cases}
\label{s:special}

Here we will discuss how we can understand the flat and the off-shell pp-wave background considered in \cite{pm08, pm0902} as special cases of the present construction. In particular, we will explain how the central charge terms arise in the Virasoro algebra and how our background independent construction explains the problem of computing Virasoro anomaly for the pp-wave as addressed in \cite{pm0902}. We will discuss the flat and the pp-wave cases separately below in subsections (\ref{ss:flat}) and (\ref{ss:pp}) respectively. Before that we make some comments based on general grounds. 

For both the backgrounds the EM tensor does not involve any non-trivial ordering between fields and conjugate momenta in the chosen coordinate system. This implies that DeWitt's generally covariant formulation is, in principle, not needed\footnote{However, we will see in subsection (\ref{ss:pp}) that understanding the pp-wave case as a special case of the background independent formulation helps us resolve certain puzzles.}. Another way of seeing this is that in both the cases we have,
\bea
\gamma_k(x) = 0 ~, \quad g(x) = 1~.
\label{gamma-g}
\eea
This indicates that any operator calculation done using the canonical commutators without worrying about manifest general covariance should be reliable. This justifies recognizing the computation done in \cite{kazama08, pm08, pm0902} as a special case of the present analysis as will be done below. It also turns out as a consequence of (\ref{gamma-g}) that the algebra in (\ref{scalar-alg}) can be considered as operator equations. 

\subsection{Flat background}
\label{ss:flat}

In this case the DeWitt-Virasoro generators are given, in the usual worldsheet language, as follows, 
\bea
\hat L^{(0)}_m = {1\over 2} \eta_{\mu \nu} \sum_{n\in Z} \hat \Pi^{\mu}_{m-n}\hat \Pi^{\nu}_n ~, 
\quad \hat{ \tilde L}^{(0)}_m = {1\over 2} \eta_{\mu \nu} \sum_{n\in Z} 
\hat{\tilde \Pi}^{\mu}_{m-n} \hat{\tilde \Pi}^{\nu}_n ~,
\label{vir-gen0}
\eea
where the superscript ${(0)}$ refers to flat background and, 
\bea
\hat \Pi^{\mu}_m &=& {1\over \sqrt{2}} \oint {d\s\over 2\pi} (\eta^{\mu \nu} \hat P_{\nu}(\s) - 
\del \hat X^{\mu}(\s)) e^{-im\s} ~, \cr
\hat{\tilde \Pi}^{\mu}_m &=& {1\over \sqrt{2}} \oint {d\s \over 2\pi} (\eta^{\mu \nu} \hat 
P_{\nu}(\s) + \del \hat X^{\mu}(\s)) e^{im\s}~,
\eea
are the usual creation-annihilation operators,
\bea
[\hat \Pi^{\mu}_m, \hat \Pi^{\nu}_n ] = \eta^{\mu \nu} \alpha' \dt_{m+n, 0} ~, \quad [\hat{\tilde \Pi}^{\mu}_m, \hat{ \tilde \Pi}^{\nu}_n ] = \eta^{\mu \nu} \alpha' \dt_{m+n, 0} ~.
\label{Pi-comm}
\eea
The DeWitt-Virasoro generators differ from the actual quantum Virasoro generators $\hat {\cal L}^{(0)}_m$ and 
${\hat{\tilde {\cal L}}}^{(0)}_m$ by additive c-numbers,
\bea
\hat L^{(0)}_m = \hat {\cal L}^{(0)}_m + c_m ~, \quad \hat{\tilde L}^{(0)}_m = \hat{\tilde {\cal L}}^{(0)}_m + \tilde c_m ~, 
\label{L-calL}
\eea
where,
\bea
\hat {\cal L}^{(0)}_m = {1\over 2} \eta_{\mu \nu} \sum_{n\in Z} :\hat \Pi^{\mu}_{m-n}\hat \Pi^{\nu}_n: ~, 
\quad \hat{ \tilde {\cal L}}^{(0)}_m = {1\over 2} \eta_{\mu \nu} \sum_{n\in Z} :\hat{\tilde 
\Pi}^{\mu}_{m-n} \hat{\tilde \Pi}^{\nu}_n: ~,
\label{vir-gen0-norm}
\eea
and $c_m = \tilde c_m$ is non-zero only when $m=0$, in which case it is a divergent constant. The normal ordering $::$ 
used in the above equations which matter only for the Virasoro zero modes is defined as the oscillator normal ordering with respect to the vacuum $|0\ra$ defined by,
\bea
\lt. \begin{array}{l} \hat \Pi^{\mu}_m \cr
\hat{\tilde \Pi}^{\mu}_m \end{array}\rt\} |0\ra = 0~, \quad \forall m\geq 0~.
\label{vac0}
\eea
It is easy to check using (\ref{Pi-comm}) that the generators in (\ref{vir-gen0}) satisfy,  
\bea
[\hat L^{(0)}_m, \hat L^{(0)}_n ] &=& (m-n)\alpha' \hat L^{(0)}_{m+n}~, \cr
[\hat{\tilde L}^{(0)}_m, \hat{\tilde L}^{(0)}_n ] &=& (m-n)\alpha' \hat{\tilde L}^{(0)}_{m+n}~, 
\cr
[\hat L^{(0)}_m, \hat{\tilde L}^{(0)}_n ] &=& 0~,
\label{vir-alg0}
\eea
which is simply the DeWitt-Virasoro algebra in (\ref{scalar-alg}) for flat background. However, the same method of computation applied to $[\hat {\cal L}^{(0)}_m, \hat {\cal L}^{(0)}_{-m}]$ gives a result that has operator ordering ambiguity. The result is $2m \alpha' \hat {\cal L}^{(0)}_0$ up to an additive c-number contribution that can not be calculated using this method because of the ambiguity. As indicated in \cite{polchinski}, this c-number contribution, which turns out to be the central charge term, can be found unambiguously by calculating, for example, $\la 0|\hat {\cal L}^{(0)}_m \hat {\cal L}^{(0)}_{-m} |0\ra $ with $m>0$,
\bea
[\hat{\cal L}^{(0)}_m, \hat{\cal L}^{(0)}_n ] &=& (m-n)\alpha' \hat{\cal L}^{(0)}_{m+n} + {D \over 12} (m^3 -m) \alpha'^2 \dt_{m+n, 0}~, \cr
[\hat{\tilde{\cal L}}^{(0)}_m, \hat{\tilde{\cal L}}^{(0)}_n ] &=& (m-n)\alpha' \hat{\tilde{\cal L}}^{(0)}_{m+n} + {D \over 12} (m^3 -m) \alpha'^2 \dt_{m+n, 0}~, 
\cr
[\hat{\cal L}^{(0)}_m, \hat{\tilde{\cal L}}^{(0)}_n ] &=& 0~.
\label{vir-alg0-norm}
\eea

\subsection{Explaining conformal invariance for pp-wave}
\label{ss:pp}

In \cite{pm0902} we considered a restricted ansatz for an off-shell pp-wave in type IIB string theory which includes the R-R plane-wave background. The R-R flux part of the background involves the Green-Schwarz fermions on the worldsheet. We will ignore this fermionic part and consider only the bosonic part of the computation which corresponds to switching on a metric-background where the non-trivial components of the metric (in physical spacetime) are given by,
\bea
G_{+-}= 1~, \quad G_{++}=K(\vec X)~, \quad G_{IJ} = \dt_{IJ}~,
\label{G-pp}
\eea
where the vector sign refers to the transverse part with index $I$. The only non-trivial component of the Ricci-tensor is,
\bea
R_{++} \propto \vec \del^2 K~.
\label{R++}
\eea
The worldsheet theory is expected to be an exact CFT when $R_{++}$ vanishes \cite{amati88}. We call the background in (\ref{G-pp}) simply as pp-wave\footnote{The particular solution of $R_{++}=0$ given by,
\bea
K = \sum_I s_I X^I X^I ~, \quad \sum_I s_I = 0~,
\eea
is called plane-wave.}. 

It was argued in \cite{pm0902} that the operator anomaly terms of the Virasoro algebra suffer from an ordering ambiguity and therefore proving conformal invariance was not completely settled. We argued below eqs.(\ref{gamma-g}) why it should be possible to consider this computation as a special case of the present background independent formulation. Here we would like to show how doing this explains the conformal invariance in the present case resolving the ambiguous situation in the previous work. 

The first step is to relate the DeWitt-Virasoro generators specialized to the present background with the quantum Virasoro generators defined in \cite{pm0902}. Given the latter, this relation is precisely the same as that in (\ref{L-calL}),
\bea
\hat L^{pp}_m = \hat {\cal L}^{pp}_m + c_m ~, \quad \hat{\tilde L}^{pp}_m = \hat{\tilde {\cal L}}^{pp}_m + \tilde c_m ~, 
\eea
where the superscript $pp$ refers to the pp-wave being considered. 
According to the calculations of the present work the algebra satisfied by $\hat L^{pp}_m$ and $\hat{\tilde L}^{pp}_m$ is given by (\ref{scalar-alg}) evaluated for the pp-wave. Following the same procedure as in flat-case, which took us from eqs.(\ref{vir-alg0}) to eqs.(\ref{vir-alg0-norm}), one finds the following quantum Virasoro algebra in the present case,
\bea
[\hat{\cal L}^{pp}_m, \hat{\cal L}^{pp}_n ] &=& (m-n)\alpha' \hat{\cal L}^{pp}_{m+n} + {D \over 12} (m^3 -m) \alpha'^2 \dt_{m+n, 0} + \hat A^R_{mn}~, \cr
[\hat{\tilde{\cal L}}^{pp}_m, \hat{\tilde{\cal L}}^{pp}_n ] &=& (m-n)\alpha' \hat{\tilde{\cal L}}^{pp}_{m+n} + {D \over 12} (m^3 -m) \alpha'^2 \dt_{m+n, 0} + \hat A^L_{mn}~, 
\cr
[\hat{\cal L}^{pp}_m, \hat{\tilde{\cal L}}^{pp}_n ] &=& \hat A_{mn}~,
\label{vir-alg-pp-norm}
\eea
where the operator anomaly terms $\hat A^R_{mn}= \hat A^R_{(i)(j)}$, $\hat A^L_{mn}=\hat A^L_{(i)(j)}$ and $\hat A_{mn}=\hat A_{(i)(j)}$ are given by eqs.(\ref{anomalies-id}) evaluated for the pp-wave. 

We will now compare the result in (\ref{vir-alg-pp-norm}) found in this work with the one in \cite{pm0902}. The computation of \cite{pm0902} was done using the local worldsheet language. The result is precisely the local version of (\ref{vir-alg-pp-norm}) with the local operator anomaly terms given by,
\bea
&& \hat A_{there}^R(\s,\s') = \hat A_{there}^L(\s,\s')= \hat A_{there}(\s,\s') \cr
&& \propto \lt[\hat {\cal O}(\s,\s') \hat P_-(\s') \hat P_-(\s') - 
\hat {\cal O}(\s',\s) \hat P_-(\s) \hat P_-(\s)\rt] \dt_{\eps}(\s-\s') \cr
&& -\lt[\hat {\cal O}(\s,\s') \del \hat X^+(\s')\del \hat X^+(\s') - \hat {\cal O}(\s',\s) \del \hat X^+(\s) \del \hat X^+(\s)\rt] \dt_{\eps}(\s-\s')~, 
\label{anomalies}
\eea
where the above three anomaly terms are defined in eqs.(3.20) of \cite{pm0902} and
\bea
\hat {\cal O}(\s, \s') = \hat P_I(\s) \del_I K(\hat{\vec X}(\s')) + 
\del_I K(\hat{\vec X}(\s')) \hat P_I(\s)~.
\eea
Therefore comparing (\ref{vir-alg-pp-norm}) with the corresponding result in \cite{pm0902} we are able to relate the operator anomaly terms in (\ref{vir-alg-pp-norm}) and in (\ref{anomalies}),
\bea
\hat A^R_{(i)(j)} &=& \oint {d\s\over 2\pi} {d\s'\over 2\pi} ~\hat A_{there}^R(\s,\s') e^{-im\s-in\s'} ~, \cr
\hat A^L_{(i)(j)} &=& \oint {d\s\over 2\pi} {d\s'\over 2\pi} ~\hat A_{there}^L(\s,\s') e^{im\s +in\s'}~, \cr 
\hat A_{(i)(j)} &=& \oint {d\s\over 2\pi} {d\s'\over 2\pi} ~\hat A_{there}(\s,\s') e^{-im\s +in\s'}~.
\label{anomalies-rel}
\eea

It was explained in \cite{pm0902} that the expression in (\ref{anomalies}) suffers from an operator ordering ambiguity. The idea here is to resolve this ambiguity by borrowing the results found here. Therefore, using eqs.(\ref{anomalies-rel}) and (\ref{anomalies-id}) one concludes that both $\hat A_{there}^R(\s,\s')$ and $\hat A_{there}^L(\s,\s')$ should vanish. Moreover, comparing the equations in (\ref{anomalies-id}), (\ref{anomalies}) and (\ref{anomalies-rel}) one concludes that the Ricci-term obtained in the present analysis was missing earlier. As we will show below, this is an apparent discrepancy which may be resolved by showing that the relevant term vanishes, though the Ricci tensor itself does not, for the pp-wave under consideration because of certain special properties of the background\footnote{This argument, however, will rely on the scalar expectation value of the algebra in (\ref{scalar-alg}), and not operator equation.}.

The non-trivial components of the infinite-dimensional metric
corresponding to (\ref{G-pp}) are,
\bea
g_{i_+j_-} = \dt_{m+n, 0}~, \quad g_{i_+j_+} = \oint {d\s \over
  2\pi}~K(\vec X(\s)) e^{i(m+n)\s}~,\quad g_{i_{\perp}j_{\perp}} =
\dt_{IJ}\dt_{m+n, 0}~, 
\label{g-pp}
\eea
where the infinite-dimensional spacetime index is divided in the
following way: $i = (i_+, i_-,  i_{\perp})$ such that,
\bea
i_+ = \{+, m\}~, \quad i_- = \{-, m\}~, \quad i_{\perp} = \{I, m\}~.
\eea
The only non-trivial components of the Ricci tensor are
$r_{i_+j_+}(\vec x)$ (the vector sign referring to the transverse
indices $i_{\perp}$) which is related to $R_{++}(\vec X)$ in (\ref{R++}) according to (\ref{rij-Rmunu}). 
Let us now go back to the last equation in (\ref{scalar-alg}). 
By going to the position space representation, integrating by parts
and using the shift property (\ref{shift}) one can show that the
Ricci-term is proportional to,
\bea 
\int dw~ \chi^* \nabla^{k+i}(r_{kl} a^{l+\bar j}) \psi = \int dx
~\chi^* g^{k_++i k'_-} \del_{k'_-}(r_{k_+l_+} a^{l_++\bar j}) \psi~,
\eea
where on the right hand side we have evaluated the term for the pp-wave. This vanishes as the quantity inside the round brackets is
independent of $x^{i_-}$.

\begin{center}
{\bf Acknowledgement}
\end{center}
I would like to thank S. R. Das, K. Dasgupta, G. Date,
J. R. David, A. Dhar, M. R. Douglas, A. R. Frey, R. Gopakumar, R. K. Kaul, 
J. Maharana, J. Maldacena, G. Mandal, S. Mukhi,
S. G. Rajeev, S. K. Rama, B. Sathiapalan , A. Sen, H. S. Sharatchandra,  
and S. R. Wadia for illuminating conversations. A preliminary
version of this work was presented at The Sixth International 
Symposium on Quantum Theory and Symmetries, University of Kentucky
in July, 2009. I thank the organizers S. R. Das and A. D. Shapere and
the participants for this stimulating conference. I am thankful to
CHEP, IISc (Bangalore),  
Department of Theoretical Physics, TIFR (Mumbai) and Department of
Physics, McGill University (Montr\'{e}al) for hospitality. Finally, I
thank M. R. Douglas and Y. Kazama for useful comments on a preliminary draft.

\appendix

\section{Some technical details regarding the map}
\label{a:map}

Here we will discuss some technical details regarding the infinite-dimensional language described 
in section \ref{s:map}. In particular we will indicate how to get the results in (\ref{GCT-g}) 
and (\ref{th-rule}).

The steps leading to the GCT property for the metric in (\ref{GCT-g}) are as follows:
\bea
g'_{ij}(x')
&=& \oint {d\s \over 2\pi}~ \Lambda_{\mu}^{~\rho}(X(\s)) \Lambda_{\nu}^{~\kappa}(X(\s)) G_{\rho 
\kappa}(X(\s)) e^{i(m+n)\s}~, \cr
&=& \sum_{r'', q'} \lt[\oint{d\s''\over 2\pi}~\Lambda_{\mu}^{~\rho}(X(\s'')) e^{-ir''\s''} \rt] 
\lt[\oint {d\s' \over 2\pi} ~\Lambda_{\nu}^{~\kappa}(X(\s'))  e^{-iq'\s'} \rt] \cr
&& \oint {d\s \over 2\pi}  ~G_{\rho \kappa}(X(\s)) e^{i(r''+m+q'+n)\s}~,\cr
&=& \sum_{r, q} \lt[\oint{d\s''\over 2\pi}~\Lambda_{\mu}^{~\rho}(X(\s'')) e^{i(m-r)\s''} \rt] 
\lt[\oint {d\s' \over 2\pi} ~\Lambda_{\nu}^{~\kappa}(X(\s'))  e^{i(n-q)\s'} \rt] \cr
&& \oint {d\s \over 2\pi}  ~G_{\rho \kappa}(X(\s)) e^{i(p+q)\s}~,\cr
&=& \lambda_i^{~l}(x) \lambda_j^{~k}(x) g_{lk}(x)~,
\eea
where in the first step we have used the transformation of the metric tensor in physical 
spacetime. To go to the second step we first write: $\displaystyle{\Lambda_{\mu}^{~\rho}(X(\s)) 
=\oint d\s'' ~\delta(\s-\s'') \Lambda_{\mu}^{~\rho}(X(\s''))}$ and 
$\displaystyle{\Lambda_{\nu}^{~\kappa}(X(\s)) =\oint d\s' ~\delta(\s-\s') 
\Lambda_{\nu}^{~\kappa}(X(\s'))}$, then we write each of the delta functions $\delta(\s-\s'')$ 
and $\delta(\s-\s')$ as infinite sum of phases as indicated in eq.(\ref{delta-0}) such that the 
summation indices are $r''$ and $q'$ respectively. Finally we rearrange the exponential factors 
suitably. In the third step we have changed the summation indices from $r''$ and $q'$ to 
$r=r''+m$ and $q=q'+n$ respectively. The tensorial property of $a^i(x)$ in eq.(\ref{GCT-g}) can 
also be established in a similar way.

To prove the general rule in (\ref{th-rule}) let us first consider $N=0$, in which case the 
general form of $u^{i_1j_1 \cdots}_{i_2j_2\cdots}(x)$ is a contraction of various factors 
involving derivatives of metric and its inverse. It can also have factors of $a^i(x)$, but we 
will consider them later. The second defining relation in (\ref{xga-XGdX}) itself is the simplest 
example of this kind. This relation ensures that,
\bea
g^{ij}(x) = \oint {d\s \over 2\pi} ~G^{\mu \nu}(X(\s)) e^{-i(m+n)\s}~,
\label{g-inv}
\eea
which also has the form (\ref{th-rule}), is the inverse metric:
\bea
g_{ij}(x) g^{jk}(x) &=& \sum_{n\in Z} \oint {d\s \over 2\pi}~ G_{\mu \nu}(X(\s)) e^{i(m+n)\s} 
\oint {d\s' \over 2\pi}~ G^{\nu \kappa}(X(\s')) e^{-i(n+q)\s'}~, \cr
&=& \oint {d\s \over 2\pi} \oint {d\s' \over 2\pi} G_{\mu \nu}(X(\s)) G^{\nu \kappa}(X(\s')) 
e^{i(m\s -q\s')} 2\pi \dt(\s-\s')~, \cr
&=& \dt_{\mu}^{\kappa} \dt_{m,q} = \dt_i^k ~,
\eea
where all the above steps are obvious. We now notice the following two properties of 
(\ref{th-rule}):
\begin{enumerate}
\item
If $v^{i\cdots}_{\cdots}(x)$ and $w^{\cdots}_{i\cdots}(x)$ satisfy (\ref{th-rule}) with 
worldsheet counterparts $V^{\mu \cdots}_{\cdots}(X(\s))$ and $W^{\cdots}_{\mu \cdots}(X(\s))$ 
respectively, then $v^{i\cdots}_{\cdots}(x)w^{\cdots}_{i\cdots}(x)$ also satisfies 
(\ref{th-rule}) with the worldsheet counterpart given by $V^{\mu \cdots}_{\cdots}(X(\s)) 
W^{\cdots}_{\mu \cdots}(X(\s))$. Here the ellipses appearing as superscripts and subscripts in 
$v$ and $V$ indicate any number of infinite-dimensional and the corresponding physical spacetime 
indices respectively.

\item
If $v^{\cdots}_{\cdots}(x)$ satisfies (\ref{th-rule}) with its worldsheet counterpart 
$V^{\cdots}_{\cdots}(X(\s))$, then $\del_i v^{\cdots}_{\cdots}(x)$ also satisfies (\ref{th-rule}) 
with the worldsheet counterpart given by
$\del_{\mu}V^{\cdots}_{\cdots}(X(\s))$. Here also the  ellipses play
similar role as before.
\end{enumerate}
The first property is easy to establish using the manipulations
leading to (\ref{g-inv}). The second one can be proved as follows: 
\bea
\del_i v^{\cdots}_{\cdots}(x) &=& \oint {d\s \over 2\pi} ~e^{im\s}
{\dt \over \dt X^{\mu}(\s)}  
\oint {d\s' \over 2\pi}~ V^{\cdots}_{\cdots}(X(\s')) e^{i(\cdots)\s'} ~, \cr
&=& \oint {d\s \over 2\pi} \oint {d\s' \over 2\pi}
~\del_{\mu}V^{\cdots}_{\cdots}(X(\s'))  
e^{im\s+i(\cdots)\s'} 2\pi \dt(\s-\s')~, \cr
&=& \oint {d\s \over 2\pi}~ \del_{\mu} V^{\cdots}_{\cdots}(X(\s))
e^{i(m+\cdots)\s}~, 
\label{del-g}
\eea
where $\cdots$ in the phase is the appropriate sum of indices that,
combined with the physical spacetime indices on $V$, give rise to the
infinite-dimensional spacetime indices on $v$. In the first two steps
of (\ref{del-g}) we have used the following results,
\bea
\del_i = \oint {d\s \over 2\pi}~ e^{im\s}{\dt \over \dt X^{\mu}(\s)}~, \quad {\dt X^{\mu}(\s) 
\over \dt X^{\nu}(\s')} = 2\pi \dt^{\mu}_{\nu} \dt(\s-\s')~.
\label{x-X}
\eea
To justify these results one may first Fourier expand: $X^{\mu}(\s)={\displaystyle \sum_{m\in Z} 
X^{\mu}_m e^{im\s}}$. Then treat the set of all string modes $\{X^{\mu}_m\}$ and the set of all 
functions $\{X^{\mu}(\s)\}$ as two sets of orthonormal and complete basis for the string 
configurations. Using the first equation in (\ref{xga-XGdX}) we identify: $x^i=X^{\mu}_m$ and 
take the orthogonality condition for these modes as:
\bea
\del_j x^i = {\del X^{\mu}_m \over \del X^{\nu}_n} = \delta^{\mu}_{\nu}\dt_{m,n}= \dt^i_j~.
\label{ortho-x}
\eea
The results in (\ref{x-X}) then follow from the above mode expansion and this orthogonality 
condition. Using the fact that the second equation of (\ref{xga-XGdX}) and (\ref{g-inv}) are of 
the form (\ref{th-rule}) and the two properties below eq.(\ref{g-inv}) one can argue that 
(\ref{th-rule}) holds true for $N=0$ if $u^{i_1j_1 \cdots}_{i_2j_2\cdots}(x)$ is independent of 
$a^i(x)$. To incorporate the $a^i(x)$ dependence we notice that the general form in which it can 
appear in $u^{i_1j_1 \cdots}_{i_2j_2\cdots}(x)$ is as follows:
\bea
u^{i_1j_1 \cdots}_{i_2j_2\cdots}(x) = v^{i_1j_1 \cdots}_{i_2j_2\cdots k_1 k_2 
\cdots}(x)a^{k_1}(x) a^{k_2}(x) \cdots ~,
\eea
where $v^{i_1j_1 \cdots}_{i_2j_2\cdots k_1 k_2 \cdots}(x)$ is independent of $a(x)$ and 
satisfies (\ref{th-rule}) with $V^{\mu_1\nu_1\cdots}_{\mu_2
\nu_2 \cdots \kappa_1 \kappa_2 \cdots} (X(\s))$ as its worldsheet counterpart.
Then using the last equation of (\ref{xga-XGdX}) one can show that $u^{i_1j_1 
\cdots}_{i_2j_2\cdots}(x)$ satisfies (\ref{th-rule}) with $V^{\mu_1\nu_1\cdots}_{\mu_2 \nu_2 
\cdots \kappa_1 \kappa_2 \cdots} (X(\s)) \del X^{\kappa_1}(\s) \del X^{\kappa_2}(\s) \cdots$ as 
its worldsheet counterpart. This proves (\ref{th-rule}) for $N=0$.

For $N\neq 0$, $u^{i_1j_1 \cdots}_{i_2j_2\cdots}(x)$ contains infinite-dimensional traces. Each 
such trace is interpreted to give a factor of $2\pi \dt(0)$ in the worldsheet language. To see 
what exactly we mean by this formal statement let us consider, for example, $v_{ijk\cdots}(x)$ 
which satisfies (\ref{th-rule}) for $N=0$ with its worldsheet counterpart $V_{\mu \nu \kappa 
\cdots}(X(\s))$. Then a single trace $u_{k\cdots}(x) \equiv g^{ij}(x)v_{ijk\cdots}(x)$ is given 
by,
\bea
u_{k\cdots }(x) &=& \sum_{m,n \in Z} \oint {d\s \over 2\pi}~G^{\mu \nu}(X(\s)) e^{-i(m+n)\s} 
\oint {d\s' \over 2\pi} ~V_{\mu \nu \kappa \cdots}(X(\s')) e^{i(m+n+q+\cdots)\s'}~, \cr
&=& \oint {d\s \over 2\pi} \oint {d\s' \over 2\pi} ~[2\pi \dt(\s-\s')]^2 G^{\mu \nu}(X(\s)) 
V_{\mu \nu \kappa \cdots}(X(\s')) e^{i(q+\cdots)\s'}~, \cr
&=& 2\pi \dt(0) \oint {d\s \over 2\pi} ~U_{\kappa \cdots}(X(\s)) e^{i(q+\cdots )\s'}~,
\eea
where $U_{\kappa \cdots}(X) = G^{\mu \nu}(X) V_{\mu \nu \kappa \cdots}(X)$.

\section{Classical Virasoro algebra in infinite-dimensional language}
\label{a:classical}

As we know, GCT is a point canonical transformation and therefore it should be possible to write 
classical Poisson brackets in a manifestly covariant form. Here we will discuss the results in 
infinite-dimensional language. Poisson bracket between any two dynamical quantities $A(x,p)$ and 
$B(x,p)$ is given by:
\bea
\{A,B\} = \sum_i \lt[{\del A\over \del x^i} {\del B\over \del p_i} - {\del B\over \del x^i} {\del 
A\over \del p_i}\rt]~.
\eea
Below we will derive the classical algebra satisfied by the generators
in  (\ref{LLtilde-classical}) and show that the known result is
obtained. The aim of this appendix is to show how the argument goes in the 
infinite-dimensional language in a manifestly covariant manner.

The algebra that we wish to compute is given by,
\bea
\{L_{(i)}, L_{(j)}\} &=& {1\over 16} \lt[T^{ZZ}_{(i)(j)} - (T^{KZ}_{(i)(j)} - i \leftrightarrow j 
) + (T^{KV}_{(i)(j)} - i \leftrightarrow j ) - (T^{ZV}_{(i)(j)} - i \leftrightarrow j ) \rt]~, \cr
\{\tilde L_{(i)}, \tilde L_{(j)}\} &=& {1\over 16} \lt[T^{ZZ}_{(\bar i)(\bar j)} + (T^{KZ}_{(\bar 
i)(\bar j)} - \bar i \leftrightarrow \bar j ) + (T^{KV}_{(\bar i)(\bar j)} - \bar i 
\leftrightarrow \bar j) + (T^{ZV}_{(\bar i)(\bar j)} - \bar i \leftrightarrow \bar j )\rt]~, \cr
\{L_{(i)}, \tilde L_{(j)}\} &=& {1\over 16} \lt[- T^{ZZ}_{(i)(\bar j)} + (T^{KZ}_{(i)(\bar j)} + 
i \leftrightarrow \bar j) + (T^{KV}_{(i)(\bar j)} - i \leftrightarrow \bar j) - (T^{ZV}_{(i)(\bar 
j)} + i \leftrightarrow \bar j )\rt]~, \cr &&
\label{vir-alg-classical-pre}
\eea
where,
\bea
T^{AB}_{(i)(j)} = \{A_{(i)}, B_{(j)}\}~.
\label{T^AB}
\eea
The terms $T^{KK}_{(i)(j)}$ and  $T^{VV}_{(i)(j)}$ have not been included in 
eqs.(\ref{vir-alg-classical-pre}) as they can be easily shown to vanish.

Let us consider the first term in eqs.(\ref{vir-alg-classical-pre}). It is straightforward to 
show:
\bea
T^{ZZ}_{(i)(j)} &=& 4 (\del_k a^{k'+i}a^{k+j}p_{k'} - i\leftrightarrow j)~.
\label{T^ZZ-pre}
\eea
This can also be written in a covariant form:
\bea
T^{ZZ}_{(i)(j)} &=& 4(\nabla_k a^{k'+i}a^{k+j}p_{k'} - i\leftrightarrow j)~,
\label{T^ZZ-int}
\eea
where $\nabla_i$ is the covariant derivative in infinite-dimensional spacetime\footnote{In worldsheet language $\nabla_i$ can be written as:
\bea
\nabla_i = \oint {d\s \over 2\pi} ~e^{im\s} \D_{X^{\mu}(\s)}~,
\eea
where $\D_{X^{\mu}(\s)}$ is the functional covariant derivative which acts on a vector, for 
example, in the following way:
\bea
\D_{X^{\nu}(\s)} V^{\mu}(X(\s')) = {\dt \over \dt X^{\nu}(\s)}V^{\mu}(X(\s')) + 2\pi \dt(\s-\s') 
\Gamma^{\mu}_{\nu \kappa}(X(\s)) V^{\kappa}(X(\s'))~,
\label{functional-cov-derv}
\eea
where $\Gamma^{\mu}_{\nu \kappa}(X)$ are the Christoffel symbols in physical spacetime whose 
expressions can be read out from those of $\gamma^i_{jk}(x)$ in (\ref{chr}) following the rules 
in (\ref{replace}). It can be explicitly checked that the derivative in 
(\ref{functional-cov-derv}) transforms covariantly under GCT:
\bea
\D_{X'^{\nu}(\s)}V'^{\mu}(X'(\s')) = \Lambda_{\nu}^{~\rho}(X(\s)) \Lambda^{\mu}_{~\s}(X(\s')) 
\D_{X^{\rho}(\s)}V^{\s}(X(\s'))~.
\eea}. The reason that the 
expressions in (\ref{T^ZZ-pre}) and (\ref{T^ZZ-int}) are same is as follows. The connection term 
coming from the covariant derivative does not contain any derivative of $a(x)$ and therefore is 
symmetric in $i$ and $j$ due to the shift property as given by the first equation in 
(\ref{shift}). Clearly this term drops out because of the anti-symmetrization in $i$ and $j$. The 
expression of $T^{ZZ}_{(i)(j)}$, however, that we would like to have for calculating the Virasoro 
algebra is as follows:
\bea
T^{ZZ}_{(i)(j)} &=& 2i(i-j)Z_{(i+j)}~,
\label{T^ZZ}
\eea
which can be obtained simply by using (\ref{del-a}) in (\ref{T^ZZ-pre}).

For the other $T^{AB}_{(i)(j)}$'s the analogues of eq.(\ref{T^ZZ-pre}) are given by:
\bea
T^{KZ}_{(i)(j)} &=& (2\del_k g^{ll'+i} a^{k+j} - 4 g^{kl+i}\del_k a^{l'+j}) p_l p_{l'}~,\cr
T^{KV}_{(i)(i)} &=& -2 g^{kl+i}\del_k (g_{k' l'}a^{k'} a^{l'+j}) p_l~, \cr
T^{ZV}_{(i)(j)} &=& -2 a^{k+i} \del_k(g_{ll'}a^l a^{l'+j})~.
\label{T-other-pre}
\eea
Using the fact that the metric is covariantly constant, such expressions can be given the 
following covariant forms:
\bea
T^{KZ}_{(i)(j)} &=& -4 g^{kl+i}\nabla_k a^{l'+j} p_l p_{l'}~,\cr
T^{KV}_{(i)(j)} &=& -4 g^{kl+i} g_{k' l'} a^{k'}\nabla_k a^{l'+j}p_l~, \cr
T^{ZV}_{(i)(j)} &=& -4 a^{k+i}g_{ll'}a^l \nabla_k a^{l'+j}~.
\label{T-other-cov}
\eea
However, we need the symmetric and anti-symmetric components of the above results to use in 
eqs.(\ref{vir-alg-classical-pre}). The anti-symmetric components are given by:
\bea
T^{KZ}_{(i)(j)} - i\leftrightarrow j &=& 4i(i-j) K_{(i+j)}~, \cr
T^{KV}_{(i)(j)} - i\leftrightarrow j &=& 2i(i-j) Z_{(i+j)}~,\cr
T^{ZV}_{(i)(j)} - i\leftrightarrow j &=& 4i(i-j)V_{(i+j)}~,
\label{anti-symm-comp}
\eea
which can be derived following the same way as (\ref{T^ZZ}) was found. The symmetric components, 
on the other hand, are as follows:
\bea
T^{KZ}_{(i)(j)} + i\leftrightarrow j &=& 0~, \cr
T^{ZV}_{(i)(j)} + i\leftrightarrow j &=& 0~.
\label{symm-comp}
\eea
To derive the first result above we first use (\ref{del-a}) and the shift symmetry (\ref{shift}) 
in the first equation of (\ref{T-other-pre}) to write:
\bea
T^{KZ}_{(i)(j)} + i\leftrightarrow j &=& 4 \sum_{l,l'} \lt[a^k \del_k g^{l+i+j l'}
-i (l+l'+i+j) g^{l+i+j l'} \rt] p_l p_{l'}~.
\label{symm-comp1}
\eea
Then we use the following identity\footnote{To derive (\ref{identity})
we first use the map (\ref{th-rule}) to write:
\bea
a^{i+k} \del_k u^{i_1j_1\cdots}_{i_2j_2\cdots}(x) &\sim & [2\pi \dt(0)]^N \oint {d\s \over 2\pi}~ 
\del X^{\mu} \del_{\mu} U^{\mu_1\nu_1\cdots}_{\mu_2\nu_2\cdots}(X(\s)) e^{i(m_2+n_2+\cdots)\s - 
i(q+m_1+n_1+\cdots)\s}~. \cr
&&
\eea
Then using $\del X^{\mu}\del_{\mu} = \del$ and finally integrating by parts one gets:
\bea
a^{i+k} \del_k u^{i_1j_1\cdots}_{i_2j_2\cdots}(x) &\sim & i\{(q+m_1+n_1+\cdots) - (m_2+n_2+\cdots 
\} \cr
&& [2\pi \dt(0)]^N \oint {d\s \over 2\pi}~ U^{\mu_1\nu_1\cdots}_{\mu_2\nu_2\cdots}(X(\s)) 
e^{i(m_2+n_2+\cdots)\s - i(q+m_1+n_1+\cdots)\s}~,
\eea
which, in the infinite-dimensional language, reads (\ref{identity}).}:
\bea
a^{i+k} \del_k u^{i_1j_1\cdots}_{i_2j_2\cdots}(x) = i\{(i+i_1+j_1+\cdots)-(i_2+j_2+\cdots)\} 
u^{i+i_1 j_1 \cdots}_{i_2j_2 \cdots}(x)~,
\label{identity}
\eea
to arrive at the first equation in (\ref{symm-comp}). Similarly, the second equation in 
(\ref{symm-comp}) can be derived by first using (\ref{del-a}) and the shift symmetry 
(\ref{shift}) in the last equation of (\ref{T-other-pre}) to arrive at:
\bea
T^{ZV}_{(i)(j)} + i\leftrightarrow j &=& -4 \sum_{l,l'} \lt[a^k \del_k g_{ll'-i-j}
+ i (l+l'-i-j) g_{l l'-i-j} \rt] a_l a_{l'}~.
\label{symm-comp2}
\eea
Then using the identity (\ref{identity}) one shows that the right hand side of the above equation 
vanishes.

Finally, one uses the results (\ref{T^ZZ}), (\ref{anti-symm-comp}) and (\ref{symm-comp}) in 
equations (\ref{vir-alg-classical-pre}) to establish the classical Virasoro algebra:
\bea
\{L_{(i)}, L_{(j)}\} &=& -i(i-j) L_{(i+j)}~, \cr
\{\tilde L_{(i)}, \tilde L_{(j)}\} &=& -i(i-j) \tilde L_{(i+j)}~, \cr
\{L_{(i)}, \tilde L_{(j)}\} &=& 0~.
\label{vir-alg-classical}
\eea

\section{Derivation of DeWitt-Virasoro algebra}
\label{a:vir-alg}

Here we will prove the results in (\ref{scalar-alg}, \ref{anomalies-id}). To proceed with the computation we will 
first introduce certain notations. The analogues of equations in (\ref{vir-alg-classical-pre})
in the present case are given by,
\bea
_{\chi}\la [\hat L_{(i)}, \hat L_{(j)}]\ra_{\psi} &=& {1\over 16} {}_{\chi}\la \hat T^{KK}_{(i)(j)} + \hat 
T^{ZZ}_{(i)(j)} - (\hat T^{KZ}_{(i)(j)} - i \leftrightarrow j ) + (\hat T^{KV}_{(i)(j)} - i 
\leftrightarrow j) \cr
&& - (\hat T^{ZV}_{(i)(j)} - i \leftrightarrow j) \ra_{\psi}~, \cr
_{\chi}\la [\hat{ \tilde L}_{(i)}, \hat{ \tilde L}_{(j)}]\ra_{\psi} &=& {1\over 16} {}_{\chi}\la \hat 
T^{KK}_{(\bar i)(\bar j)} + \hat T^{ZZ}_{(\bar i)(\bar j)} + (\hat T^{KZ}_{(\bar i)(\bar j)} - 
\bar i \leftrightarrow \bar j) + (\hat T^{KV}_{(\bar i)(\bar j)} - \bar i \leftrightarrow \bar j) 
\cr
&& + (\hat T^{ZV}_{(\bar i)(\bar j)} - \bar i \leftrightarrow \bar j ) \ra_{\psi}~, \cr
_{\chi}\la[\hat L_{(i)}, \hat{\tilde L}_{(j)}]\ra_{\psi} &=&{1\over 16}{}_{\chi}\la \hat T^{KK}_{(i)(\bar 
j)} - \hat T^{ZZ}_{(i)(\bar j)} + (\hat T^{KZ}_{(i)(\bar j)} + i \leftrightarrow \bar j) + (\hat 
T^{KV}_{(i)(\bar j)} - i \leftrightarrow \bar j) \cr
&& - (\hat T^{ZV}_{(i)(\bar j)} + i \leftrightarrow \bar j )\ra_{\psi}~, \cr &&
\label{vir-alg-quantum-pre}
\eea
where $_{\chi}\la \cdots \ra_{\psi} = \la \chi |\cdots |\psi \ra$, $|\chi \ra$ and $|\psi \ra$ 
being two arbitrary spin zero states. The operator $\hat T^{AB}_{(i)(j)}$ is given by the quantum 
version of (\ref{T^AB}):
\bea
\hat T^{AB}_{(i)(j)} = [\hat A_{(i)}, \hat B_{(j)}]~,
\eea
where $\hat K_{(i)}$, $\hat Z_{(i)}$ and $\hat V_{(i)}$ are defined in equations 
(\ref{KZV-quantum}).

Below we will compute the various expectation values that appear on the right hand sides of 
eqs.(\ref{vir-alg-quantum-pre}). Such calculations are done by using the following basic results 
extensively:
\bea
{}_{\chi}\la \hat K_{(i)}\ra_x &=& -\alpha'^2 \nabla_{(i)}^2 \chi^*(x) ~,
\label{chiKx} \\
{}_x\la \hat K_{(i)}\ra_{\psi} &=& -\alpha'^2 \nabla_{(i)}^2 \psi(x)~,
\label{xKpsi} \\
{}_{\chi}\la\hat Z^L_{(i)}\ra_x &=& i\alpha' \nabla_k \chi^*(x) a^{k+i}(x)~,
\label{chiZLx} \\
{}_x\la \hat Z^L_{(i)}\ra_{\psi} &=& -i\alpha' \nabla_k \lt(a^{k+i}(x) \psi(x)\rt)~,
\label{xZLpsi} \\
{}_{\chi}\la \hat Z^R_{(i)}\ra_x &=& i\alpha' \nabla_k\lt( \chi^*(x) a^{k+i}(x) \rt)~,
\label{chiZRx} \\
{}_x\la \hat Z^R_{(i)}\ra_{\psi} &=& -i\alpha' a^{k+i}(x) \nabla_k \psi(x)~,
\label{xZRpsi}
\eea
where $\nabla_{(i)}^2 = g^{k+i k'}(x) \nabla_k \nabla_{k'}$ is the {\it shifted} Laplace-Beltrami 
operator. These equations can be easily derived by using the basic definitions in (\ref{pih-def}) 
and (\ref{p-rep}).

Let us first consider ${}_{\chi}\la \hat T^{KK}_{(i)(j)}\ra_{\psi}$. Using (\ref{chiKx}) and 
(\ref{xKpsi}) one finds:
\bea
{}_{\chi}\la \hat K_{(i)}\hat K_{(j)}\ra_{\psi}  &=& \alpha'^4  \int dw~ \nabla_{(i)}^2 \chi^* 
\nabla_{(j)}^2 \psi = \alpha'^4  \int dw~ \chi^* \nabla_{(i)}^2 \nabla_{(j)}^2 \psi~,
\eea
where in the last step we have used integrations by parts to move the derivatives from $\chi^*$ 
to $\psi$. Using the shift property in (\ref{shift}) one concludes right away that the above 
expression is symmetric in $i$ and $j$ and therefore the commutator ${}_{\chi}\la \hat 
T^{KK}_{(i)(j)}\ra_{\psi}$ must vanish. However, we would like to compute it by explicitly 
calculating the commutator of the shifted Laplace-Beltrami operators without imposing the shift 
property until the very end. This yields:
\bea
{}_{\chi}\la \hat T^{KK}_{(i)(j)}\ra_{\psi}
&=& \alpha'^4 \int dw~ \lt(\nabla^{k+i} \chi^* r_{kl} \nabla^{l+j} \psi - \nabla^{k+j} \chi^* 
r_{kl} \nabla^{l+i} \psi \rt)~, \cr
&=& \alpha'^2 {}_{\chi} \la \hat \pi^{\star k+i} r_{kl}(\hat x) \hat \pi^{l+j} - i 
\leftrightarrow j \ra_{\psi} ~, \cr
&=& 0~,
\label{T^KK}
\eea
where we have used the fact that commutator of covariant derivatives acting on a scalar vanishes, 
but yields the following for a vector $V$,
\bea
[\nabla_i, \nabla_j] V^k = r^k_{lij}V^l~,
\eea
where $r^k_{lij}$ is the Riemann tensor of the infinite-dimensional spacetime. Notice the 
appearance of the Ricci tensor $r_{ij}=r^k_{ikj}$ in (\ref{T^KK}).
We will find such contributions many times in the rest of the computations.
Such terms drop out of the final result in the present computation because of the 
anti-symmetrization in $i$ and $j$. However, as we will see, in certain other computations they 
will survive.

We will now consider the following term:
\bea
{}_{\chi}\la \hat T^{ZZ}_{(i)(j)} \ra_{\psi}&=& {}_{\chi}\la [\hat T^{Z^LZ^L}_{(i)(j)} + \hat 
T^{Z^RZ^R}_{(i)(j)} + (\hat T^{Z^LZ^R}_{(i)(j)} - i\leftrightarrow j)]\ra_{\psi}~.
\label{chiT^ZZpsi-pre}
\eea
Using (\ref{chiZLx}) and (\ref{xZLpsi}) one may write:
\bea
{}_{\chi}\la \hat Z^L_{(i)} \hat Z^L_{(j)} \ra_{\psi} &=& \alpha'^2 \int dw~ \nabla_k \chi^* 
a^{k+i} \nabla_l \lt( a^{l+j}\psi \rt)~, \cr
&=& -\alpha'^2 \int dw~ \lt(\nabla_k \nabla_l \chi^* a^{k+i} a^{l+j}\psi + \nabla_k \chi^* 
\nabla_l a^{k+i} a^{l+j} \psi \rt) ~.
\eea
Anti-symmetrizing the above expression and using (\ref{del-a}) and the shift property one finally 
finds,
\bea
{}_{\chi}\la \hat T^{Z^LZ^L}_{(i)(j)} \ra_{\psi}
&=& -(i-j) \alpha' ~{}_{\chi}\la \hat Z^L_{(i+j)} \ra_{\psi}~.
\label{chiZLZLpsi}
\eea
Similarly,
\bea
{}_{\chi}\la \hat T^{Z^RZ^R}_{(i)(j)}\ra_{\psi}
&=& -(i-j) \alpha' ~{}_{\chi}\la \hat Z^R_{(i+j)} \ra_{\psi}~.
\label{chiZRZRpsi}
\eea
To compute the last term in (\ref{chiT^ZZpsi-pre}) we first derive:
\bea
{}_{\chi}\la \hat T^{Z^LZ^R}_{(i)(j)}\ra_{\psi} &=& -\alpha'^2 \int dw~ \lt[\chi^* \lt(\nabla_l 
a^{k+i} \nabla_k a^{l+j} + a^{k+i} r_{kl} a^{l+j} \rt)\psi \rt. \cr
&& \lt. + \nabla_k \chi^* a^{l+j} \nabla_l a^{k+i} \psi + \chi^* \nabla_k a^{l+j} a^{k+i} 
\nabla_l \psi \rt]~.
\eea
Terms in the first line drop off when we anti-symmetrize between $i$ and $j$. The final result 
can be written in the following form:
\bea
{}_{\chi}\la \hat T^{Z^LZ^R}_{(i)(j)}\ra_{\psi} - i\leftrightarrow j &=&
-(i-j) \alpha' ~{}_{\chi}\la \hat Z_{(i+j)}\ra_{\psi}
-\alpha'^2 \lt({}_{\chi}\la a^{k+i}(\hat x)r_{kl}(\hat x) a^{l+j}(\hat x)\ra_{\psi} - 
i\leftrightarrow j \rt)
~, \cr
&=& -(i-j) \alpha' ~{}_{\chi}\la \hat Z_{(i+j)} \ra_{\psi} ~,
\label{chiZLZRpsi-anti}
\eea
where again the terms involving the Ricci tensor cancel.
Accumulating all the results in (\ref{chiZLZLpsi}), (\ref{chiZRZRpsi}) and 
(\ref{chiZLZRpsi-anti}) one finally gets:
\bea
{}_{\chi} \la \hat T^{ZZ}_{(i)(j)} \ra_{\psi} &=& -2(i-j)\alpha' ~{}_{\chi}\la \hat 
Z_{(i+j)}\ra_{\psi} -\alpha'^2 \lt({}_{\chi}\la a^{k+i}(\hat x)r_{kl}(\hat x)a^{l+j}(\hat 
x)\ra_{\psi} - i\leftrightarrow j \rt)~,  \cr
&=& -2(i-j)\alpha' ~{}_{\chi}\la \hat Z_{(i+j)}\ra_{\psi}~.
\label{chiT^ZZpsi}
\eea

Next we proceed to compute the symmetric and anti-symmetric parts of $\hat T^{KZ}_{(i)(j)}$. 
Following similar manipulations as above we first arrive at:
\bea
{}_{\chi}\la \hat T^{KZ^L}_{(i)(j)}\ra_{\psi} &=& i\alpha'^3 \int dw~ \lt[\nabla^{k+i}\chi^* 
r_{kl} a^{l+j} \psi +\nabla^{l+i}\nabla_k \chi^* \nabla_l a^{k+j} \psi \rt.\cr
&& \lt. - \nabla_k \chi^*\nabla_l a^{k+j} \nabla^{l+i}\psi \rt]~,
\label{chiKZLpsi-comm1}
\eea
Then we manipulate the last two terms in a certain way. Evaluating the covariant derivative on 
$a$ in the second term one gets:
\bea
\nabla^{l+i}\nabla_k \chi^* \nabla_l a^{k+j} \psi &=& i\sum_k (k+j) \nabla_k \nabla^{k+i+j} 
\chi^* \psi + \gamma^k_{ll'} \nabla^{l+i+j}\nabla_k \chi^* a^{l'}\psi ~,
\label{second-term}
\eea
where we have written the summation symbol explicitly in the first term because of the $k$ 
dependent pre-factor. The connection term is manipulated in the following way:
\bea
\gamma^k_{ll'} \nabla^{l+i+j}\nabla_k \chi^* a^{l'}\psi
&=& {1\over 2} \lt(\gamma^{k+i+j}_{l' \tilde k}g^{\tilde k l} + \gamma^{l+i+j}_{l'\tilde l} g^{k 
\tilde l}\rt) \nabla_k \nabla_l \chi^* a^{l'}\psi ~, \cr
&& \lt[\hbox{Using shift property (\ref{shift}) } \rt] \cr
&& \cr
&=& - {1\over 2} \del_{l'} g^{k l+i+j}\nabla_k \nabla_l \chi^* a^{l'} \psi ~, \cr
&& \lt[\hbox{Using: } \nabla_{l'}g^{kl} = \del_{l'} g^{kl} + \gamma^k_{l' \tilde k}g^{\tilde kl} 
+ \gamma^l_{l'\tilde l} g^{k\tilde l} = 0~. \rt] \cr &&\cr
&=& - {i\over 2}\sum_{k,l}(k+l+i+j) g^{k+i+j l} \nabla_k \nabla_l \chi^* \psi ~,\cr
&& [\hbox{Using (\ref{identity}).}] \cr && \cr
&=& -i \sum_k (k+{i+j\over 2}) \nabla_k \nabla^{k+i+j} \chi^* \psi ~,
\label{connection-term}
\eea
Substituting the result (\ref{connection-term}) in (\ref{second-term}) one writes for the second 
term in (\ref{chiKZLpsi-comm1}):
\bea
\int dw~ \nabla^{l+i}\nabla_k \chi^* \nabla_l a^{k+j} \psi &=& {i\over 2} (j-i) \int dw~ 
\nabla_{(i+j)}^2 \chi^* \psi ~. \cr &&
\label{KZL-comm-term2}
\eea
Substituting this results into (\ref{chiKZLpsi-comm1}) and evaluating the covariant derivative on 
$a$ in the last term of the same one finally gets:
\bea
{}_{\chi}\la \hat T^{KZ^L}_{(i)(j)}\ra_{\psi}
&=& i\alpha'^3 \int dw~ \lt[\nabla^{k+i}\chi^* r_{kl} a^{l+j} \psi + i\sum_k(k+{3j\over 2}-{i\over 2}) 
\nabla_k \nabla^{k+i+j} \chi^* \psi  \rt. \cr
&& \lt. - a^{l'} \gamma^{k+i+j}_{l'l} \nabla_k \chi^* \nabla^l \psi \rt]~.
\label{chiKZLpsi-comm2}
\eea
A similar manipulation gives the following result:
\bea
{}_{\chi}\la \hat T^{KZ^R}_{(i)(j)}\ra_{\psi}
&=& i\alpha'^3 \int dw~ \lt[\chi^* a^{k+i}r_{kl}\nabla^{l+j}\psi +
i\sum_k (k+{3j\over 2}-{i\over 2}) 
\nabla_k \nabla^{k+i+j} \chi^* \psi  \rt. \cr
&& \lt. - a^{l'}\gamma^{k+i+j}_{l'l} \nabla_k \psi \nabla^l \chi^* \rt]~.
\label{chiKZRpsi-comm}
\eea
Combining the results (\ref{chiKZLpsi-comm2}) and (\ref{chiKZRpsi-comm}) we get,
\bea
{}_{\chi}\la \hat T^{KZ}_{(i)(j)}\ra_{\psi}
&=& i\alpha'^3 \int dw~ \lt[\nabla^{k+i}\chi^* r_{kl} a^{l+j} \psi + \chi^* 
a^{k+i}r_{kl}\nabla^{l+j}\psi \rt. \cr
&& \lt. + i\sum_k (2k+3j-i) \nabla_k \nabla^{k+i+j} \chi^* \psi
- a^{l'} \gamma^{k+i+j}_{l'l} \lt(\nabla_k \chi^* \nabla^l \psi + \nabla_k \psi \nabla^l \chi^* 
\rt)\rt] ~. \cr
&&
\label{chiKZpsi-comm1}
\eea
We further manipulate the last term in the following way:
\bea
a^{l'} \gamma^{k+i+j}_{l'l} \lt(\nabla_k \chi^* \nabla^l \psi + \nabla_k \psi \nabla^l \chi^* \rt)
&=& a^{l'} \lt(\gamma^{k+i+j}_{l'\tilde k} g^{\tilde k l} + \gamma^{l+i+j}_{l' \tilde l} 
g^{\tilde l k} \rt) \nabla_k \chi^* \nabla_l \psi ~, \cr
&=& -a^{l'} \del_{l'} g^{k+i+j l} \nabla_k \chi^* \nabla_l \psi~, \cr
&=& -i\sum_k (2k+i+j) \nabla_k\chi^* \nabla^{k+i+j} \psi~.
\eea
Substituting this result into (\ref{chiKZpsi-comm1}) one finally obtains:
\bea
{}_{\chi}\la \hat T^{KZ}_{(i)(j)}\ra_{\psi}
&=& -2\alpha' (i-j) {}_{\chi}\la \hat K_{(i+j)} \ra_{\psi} \cr
&& + \alpha'^2 {}_{\chi}\la \hat \pi^{\star k+i} r_{kl}(\hat x) a^{l+j}(\hat x) - a^{k+i}(\hat x) 
r_{kl}(\hat x) \hat \pi^{l+j} \ra_{\psi}~.
\label{chiKZpsi-comm2}
\eea
Notice that the first term is anti-symmetric in $i$ and $j$, whereas the rest is symmetric 
because of the shift property. Therefore,
\bea
{}_{\chi}\la \hat T^{KZ}_{(i)(j)}\ra_{\psi} - i\leftrightarrow j &=& -4\alpha' (i-j) {}_{\chi}\la 
\hat K_{(i+j)} \ra_{\psi}~,
\label{T^KZ-anti} \\
{}_{\chi}\la \hat T^{KZ}_{(i)(j)}\ra_{\psi} + i\leftrightarrow j &=& 2\alpha'^2 {}_{\chi}\la \hat 
\pi^{\star k+i} r_{kl}(\hat x) a^{l+j}(\hat x) - a^{k+i}(\hat x) r_{kl}(\hat x) \hat \pi^{l+j} 
\ra_{\psi}~. \cr &&
\label{T^KZ-symm}
\eea

The anti-symmetric part of ${}_{\chi}\la \hat T^{KV}_{(i)(j)}\ra_{\psi}$ is straightforward to 
compute leading to the following result:
\bea
{}_{\chi}\la \hat T^{KV}_{(i)(j)} \ra_{\psi} - i\leftrightarrow j
&=& -2(i-j)\alpha' ~{}_{\chi}\la \hat Z_{(i+j)}\ra_{\psi}~.
\label{T^KV}
\eea

Finally, we need to compute ${}_{\chi}\la \hat T^{ZV}_{(i)(j)}\ra_{\psi}$. We will establish the 
following relations:
\bea
{}_{\chi}\la \hat T^{Z^LV}_{(i)(j)}\ra_{\psi} = {}_{\chi}\la \hat T^{Z^RV}_{(i)(j)}\ra_{\psi} = 
-(i-j)\alpha' ~{}_{\chi}\la\hat V_{(i+j)}\ra_{\psi}~,
\label{T^ZLV-T^ZRV}
\eea
such that we have the following results for the desired quantities:
\bea
{}_{\chi}\la \hat T^{ZV}_{(i)(j)}\ra_{\psi} -i\leftrightarrow j &=& -4(i-j)\alpha' ~{}_{\chi}\la\hat 
V_{(i+j)}\ra_{\psi}~,\cr
{}_{\chi}\la \hat T^{ZV}_{(i)(j)}\ra_{\psi} +i\leftrightarrow j &=& 0~.
\label{T^ZV}
\eea
In order to establish (\ref{T^ZLV-T^ZRV}) let us consider, for example, ${}_{\chi}\la 
\hat T^{Z^LV}_{(i)(j)}\ra_{\psi}$. It is straightforward to show:
\bea
{}_{\chi}\la \hat T^{Z^LV}_{(i)(j)}\ra_{\psi} &=& -2i\alpha' \int dw~ \chi^* a^{k+i} g_{ll'}\nabla_k 
a^l a^{l'+j} \psi~, \cr
&=& -2i\alpha' \int dw~ \lt[i\sum_k(k)\chi^*g_{kk'}a^{k+i}a^{k'+j}\psi + a^{\tilde 
k}\gamma^l_{\tilde kk}g_{ll'} a^{k+i}a^{l'+j}\chi^*\psi \rt] ~,
\label{T^ZLV-pre}
\eea
where in the second step we have evaluated the covariant derivative. We manipulate the connection 
term as follows:
\bea
a^{\tilde k}\gamma^l_{\tilde kk}g_{ll'} a^{k+i}a^{l'+j}\chi^*\psi &=& {1\over 2} a^{\tilde 
k}\lt(\gamma^l_{\tilde kk}g_{ll'} +\gamma^l_{\tilde k l'}g_{lk}\rt) a^{k+i}a^{l'+j}\chi^*\psi  ~, 
\cr
&&\lt[\hbox{Using: } a^{k+i}a^{l'+j} = a^{l'+i}a^{k+j}~. \rt] \cr && \cr
&=& {1\over 2} a^{\tilde k} \del_{\tilde k} g_{kl'} a^{k+i}a^{l'+j}\chi^*\psi ~, \cr
&& \lt[\hbox{Using: } \nabla_{\tilde k}g_{kl'}= \del_{\tilde k} g_{kl'}- \lt(\gamma^l_{\tilde 
kk}g_{ll'} +\gamma^l_{\tilde k l'}g_{lk}\rt) = 0~.\rt]~, \cr && \cr
&=& - {i\over 2} \sum_{k,k'}(k+k') \chi^*g_{kk'} a^{k+i}a^{k'+j}\psi ~,
\eea
where in the last step we have used (\ref{identity}). Using the above result in (\ref{T^ZLV-pre}) 
one gets:
\bea
{}_{\chi}\la \hat T^{Z^LV}_{(i)(j)}\ra_{\psi} &=& \sum_{k,k'} (k-k') \alpha' \int 
dw~\chi^*g_{kk'}a^{k+i}a^{k'+j}\psi ~, \cr
&=& -(i-j)\alpha'\int dw~ \chi^* g_{k-i k'-j} a^k a^{k'} \psi + \sum
_{k,k'} (k-k')\alpha' \int dw~ \chi^* g_{k-i k'-j} a^k a^{k'} \psi~,\cr &&
\eea
where in the second step we have shifted the summation variables $k\to
k-i$ and $k'\to k'-j$. The second term in the last equation vanishes
as the integrand is symmetric under $k \leftrightarrow k'$. The first
term gives the desired result. The result for ${}_{\chi}\la \hat T^{Z^RV}_{(i)(j)}\ra_{\psi}$ in 
(\ref{T^ZLV-T^ZRV}) can also be established using similar arguments.

Substituting the results (\ref{T^KK}, \ref{T^ZZ}, \ref{T^KZ-anti}, \ref{T^KZ-symm}, \ref{T^KV}, 
\ref{T^ZV}) in (\ref{vir-alg-quantum-pre}) one finally establishes the results (\ref{scalar-alg}).

\end{document}